\pdfoutput=1
\documentclass{egpubl}
\usepackage{eg2020}

\ConferencePaper        
\usepackage[T1]{fontenc}
\usepackage{dfadobe}

\usepackage{cite}  
\BibtexOrBiblatex
\electronicVersion
\PrintedOrElectronic
\ifpdf \usepackage[pdftex]{graphicx} \pdfcompresslevel=9
\else \usepackage[dvips]{graphicx} \fi

\usepackage{egweblnk}

\usepackage{subcaption}
\usepackage{tabularx}
\usepackage{bm}
\usepackage{multirow}

\title[Organic Narrative Charts]%
      {Organic Narrative Charts}

\author[F. Bolte \& S. Bruckner]
{\parbox{\textwidth}{\centering
        F. Bolte\orcid{0000-0003-4472-5380}
        and S.Bruckner\orcid{0000-0002-0885-8402}
        }
        \\
{\parbox{\textwidth}{\centering University of Bergen, Norway}
}
}

\begin{document}


\maketitle
\begin{abstract}
Storyline visualizations display the interactions of groups and entities and their development over time. Existing approaches have successfully adopted the general layout from hand-drawn illustrations to automatically create similar depictions. Ward Shelley is the author of several diagrammatic paintings that show the timeline of art-related subjects, such as \textit{Downtown Body}, a history of art scenes. His drawings include many stylistic elements that are not covered by existing storyline visualizations, like links between entities, splits and merges of streams, and tags or labels to describe the individual elements. We present a visualization method that provides a visual mapping for the complex relationships in the data, creates a layout for their display, and adopts a similar styling of elements to imitate the artistic appeal of such illustrations. We compare our results to the original drawings and provide an open-source authoring tool prototype.


\end{abstract}

\section{Introduction}\label{sec:introduction}

Movie Narrative Charts~\cite{munroe2009movie} are hand-drawn illustrations that display characters as streams and show their interactions in different locations over time. They have inspired a multitude of works in the visualization community and several attempts have been made to create digital replicas. In the same manner, we were inspired by Ward Shelley's diagrammatic paintings, which cover complex relationships and a multitude of textual annotations to display the evolution of art related subjects. We aim to create a digital chart that captures the organic appearance, displays the complex relationships between individual entities, and imitates the artistic appeal of these drawings. The digital support for such diagrams would not only allow for the fast creation of visually appealing results from varying data sources, but further ease the lengthy planning process that artists undergo to create a single piece of work. The provision of an authoring tool for artists could allow them to focus more on the artistic part, such as the creation of a general theme and bigger picture, that automatic systems will not be able to create.

We contribute a more organic layout for narrative charts that combines nesting, splits, merges, parent changes, and multiple label types. We apply image processing techniques to create a visual appeal similar to hand-drawn paintings by artists, while still complying with the SVG standard for further editing. Our prototype of an interactive authoring tool is made publicly available.

\begin{figure}
	\centering{\includegraphics[width=\columnwidth]{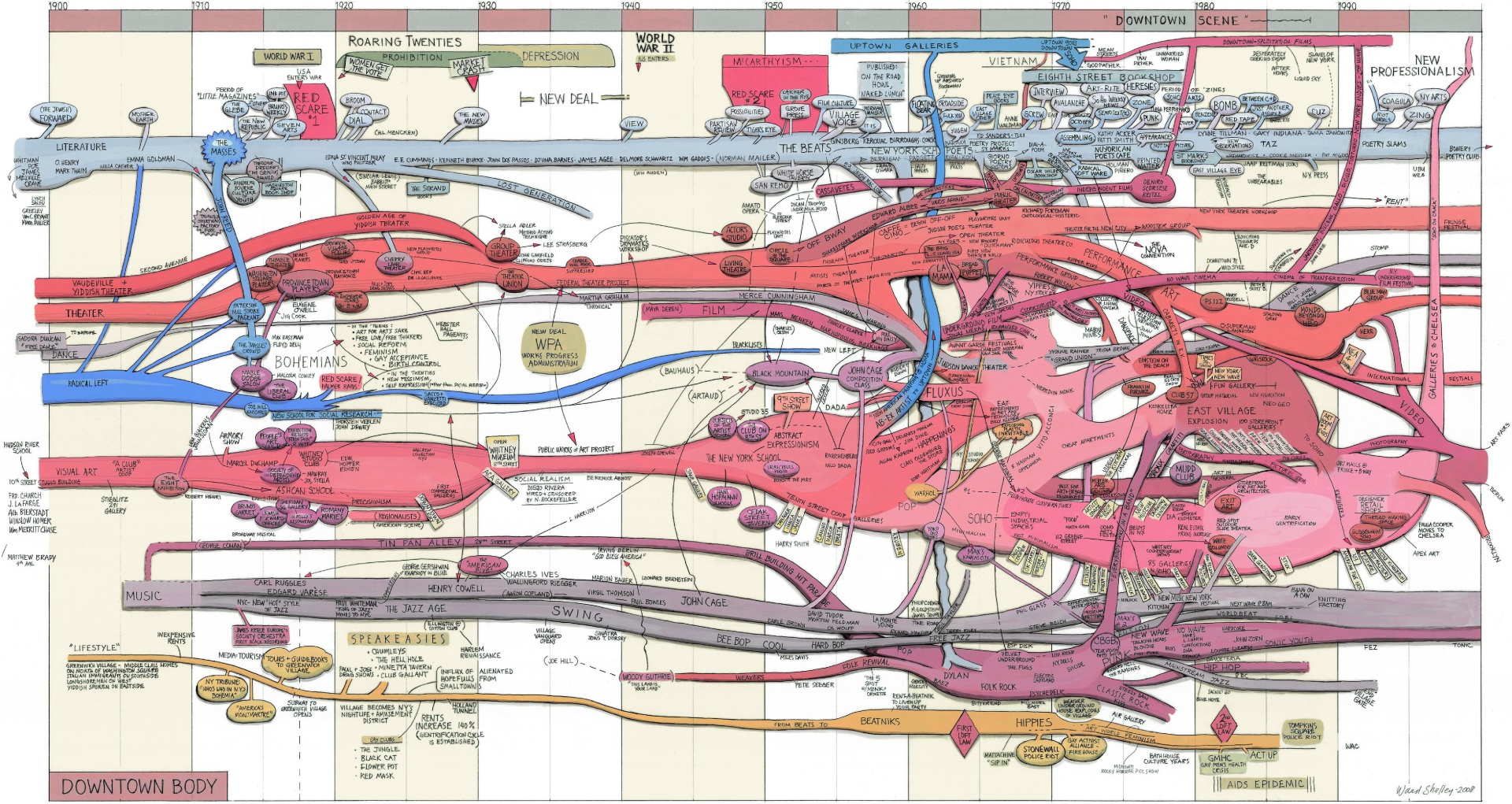}}
	\caption{Downtown Body by Ward Shelley~\cite{wardshelley}}
	\label{fig:downtown}
\end{figure}

\section{Related Work}\label{sec:relatedwork}

While a lot of research deals with time-related data, layouting techniques, and arts, we put our focus on stream-based visualizations.

Ogawa and Ma~\cite{ogawa2010software}  automated the creation of narrative charts, whose aesthetics were improved by Tanahashi and Ma~\cite{tanahashi2012design}. StoryFlow~\cite{liu2013storyflow} applies an adaption of a layered graph layout~\cite{sugiyama1981methods} to improve the performance of the aesthetics optimization.
Finally, Tang et al.~\cite{tang2018istoryline} analyzed the differences between automatic layout approaches and artist desires and developed iStoryline as an authoring tool for the creation of storyline visualizations.
In narrative charts, streams move together in groups, but can not be individually divided or merged. In our case, on the other hand, we are looking at streams with changing height values that feature direct connections, merges, splits, and a predominant amount of text labels.

TextFlow~\cite{cui2011textflow} and Xu et al.~\cite{xu2013visual} display topic splits and merges based on the ThemeRiver metaphor~\cite{havre2002themeriver}, facilitate a force layout to generate aesthetically pleasing results, and integrate word clouds into flow charts. Our approach adds an artistic style closer to hand-drawn charts, supports a larger amount of text labels directly integrated into the layout algorithm, and provides interactions for an authoring tool prototype.

\section{Overview}\label{sec:overview}

Ward Shelley's diagrammatic paintings typically describe the evolution of a given topic over time. \textit{Downtown Body} (\autoref{fig:downtown}), for example, shows how different branches of art, like theater, music, and visual arts, changed and interacted throughout the 20th century. His drawings not only provide a general overview of topics and their relevance over the years, but further highlight individual milestones of every time period.
By analyzing his work we identified three elements that define the main structure of his paintings. We will in the following refer to them as \textit{streams}, \textit{labels}, and \textit{links}. We further analyzed the artistic \textit{style} of the paintings to take inspiration for creating similar digital results.

\noindent\textbf{Streams:} All elements in Shelley's paintings follow a common timeline. A stream is defined for a certain period of time and describes one of the main topics of interest. For example, one stream might represent \textit{Literature} from 1898 until the year 2000, while another stream called \textit{Film} only starts to show in the year 1942. Streams are represented by colored lines whose thickness can change over time to convey the changing relevance of this topic~(\autoref{fig:overview} A). In addition to the main topics, there might be relevant subtopics that the artist wants to highlight. The \textit{New York School} and \textit{Early Gentrification} in \textit{Visual Arts} can be seen as such an example. They are typically represented by a stream of slightly different shade than the main topic, nested inside the main stream~(\autoref{fig:overview} C).

Streams and their represented topics can merge and split at specific points in time. When multiple streams merge, they will be represented by a single line after the given timepoint. In the same manner, a splitting stream will be drawn as multiple lines after the given point in time and represent multiple topics.

\noindent\textbf{Labels:} One of the major elements that contributes to the vast amount of information these drawings contain is text. Every stream is labeled by its represented topic when it appears. Additional text labels are describing the narrative of the stream by naming major events or core contributions to that field. In the \textit{Literature} topic, such labels include authors like \textit{Emma Goldman}, general movements like the Beat Generation (\textit{The Beats}), newsletters, book stores, and events like \textit{Poetry Slams}. The majority of labels is connected to at least one of the topic streams and referring to a certain point or period in time.

Labels come in many different forms and shapes, but can be classified into three major classes: \textit{outside}, \textit{inside}, and \textit{on top} of streams. An example for each type can be found in \autoref{fig:overview}. Outside labels are typically surrounded by another shape that is connected to the referred to stream through a line in the same color. Inside labels have a shape similar to outside labels, but are missing the connecting line. Instead, they touch the stream they refer to, or are drawn inside it. Their background color is different to the color of the stream, but often similar in hue. A label drawn on top of streams is not surrounded by a shape, but directly drawn onto the stream itself. Its text bends along the overall stream shape.  

All text labels are black with all letters capitalized. Their font sizes can vary to emphasise on the importance of individual elements. While inside and outside labels can feature arbitrary surrounding shapes, they are mainly surrounded by ellipses and rectangles. By utilizing unique shapes for specific labels they can stand out from the rest.

\noindent\textbf{Links:} Links communicate the connection of two entities, joining streams with streams, labels with labels, labels with streams, and even other links to streams or labels. They are defined by a start and an end point, each of which is defined by a point in time and a connected entity. If the entities at the start and end point feature different colors, the link is typically drawn in the color of the start entity. Links can further be drawn as thin black lines or arrows. Compared to streams, links typically do not change in size and rarely contain labels. Furthermore, while streams typically exist throughout a larger time period, links are rather short-lived, which leads to a general difference in appearance. While streams follow the time axis of the diagram, links are typically almost perpendicular to it.

\noindent\textbf{Style:} Ward Shelley's paintings feature an iconic time axis where time is discretized into blocks of alternating color. These blocks are drawn on the time axis itself, as well as in the background of the painting, where they can use different colors and a different discretization. Colors on the time axis are generally more saturated.

All shapes are typically filled by a solid color categorizing the topic of the element. Nested streams utilize a different shade of the same color to appear similar. The elements feature a strong black outline, as well as an inner and outer shadow towards the bottom left. The general shape of elements appears organic and follows a common scheme to create a bigger picture like blood vessels in the body (\textit{Downtown Body}), or a tentacled beast (\textit{History of Science Fiction}). This reoccurring schematic style inspired the name for our organic narrative charts. In some of the drawings we found additional complimentary embellishments like pictures of human bodies (\textit{Carolee Schneemann}) or album covers (\textit{Frank Zappa}).

\section{Method}\label{sec:method}

\begin{figure}[ht]
    \centering
    \begin{subfigure}{\columnwidth}
        \centering
        
    
    \setlength{\tabcolsep}{7pt} 
    \renewcommand{\arraystretch}{0.85}
    \small
    \begin{tabular}{l|cccccc}\hline
        \multirow{4}{*}{Streams} & \textbf{ID} & \textbf{$t_0$} & \textbf{$t_1$} & \textbf{color} & \textbf{size} & \textbf{parent} \\ \cline{2-7} 
        &A & 2& 6& \#D73   &      &  \\ 
        &B & 3& 9& blue    & 5/10 &  \\ 
        &C & 4& 6& purple &      & B \\ \hline
        
        \multirow{3}{*}{Links} &\multicolumn{2}{c}{\textbf{from}} & \textbf{$t_0$}& \textbf{to}& \textbf{$t_1$} & \textbf{merge} \\ \cline{2-7}
        & \multicolumn{2}{c}{\textbf{A}} & 3 & B & & true \\ 
         &\multicolumn{2}{c}{\textbf{C}} & 4 & A & &  \\ \hline
         
        \multirow{4}{*}{Lables} &\multicolumn{2}{c}{\textbf{stream}} & \textbf{t} & \textbf{text} & \textbf{type} & \textbf{size(em)}\\ \cline{2-7}
        &\multicolumn{2}{c}{\textbf{A}} & 4 & inside label   & in  & 3\\ 
        &\multicolumn{2}{c}{\textbf{B}} & 6 & outside ...    & out & 5\\ 
        &\multicolumn{2}{c}{\textbf{B}} & 7 & ... on top ... & on  & 3\\ \hline

    \end{tabular}
        \caption{CSV data. The specified data format enables a simple definition of all diagram elements. Each stream has an ID and is defined in a given time period. Size is a list of tuples, where the first element refers to a point in time and the second to the size. A parent is only defined for nested streams.}
        \label{tab:data}
    \end{subfigure}
    
    \begin{subfigure}{\columnwidth}
        \includegraphics[width=\linewidth]{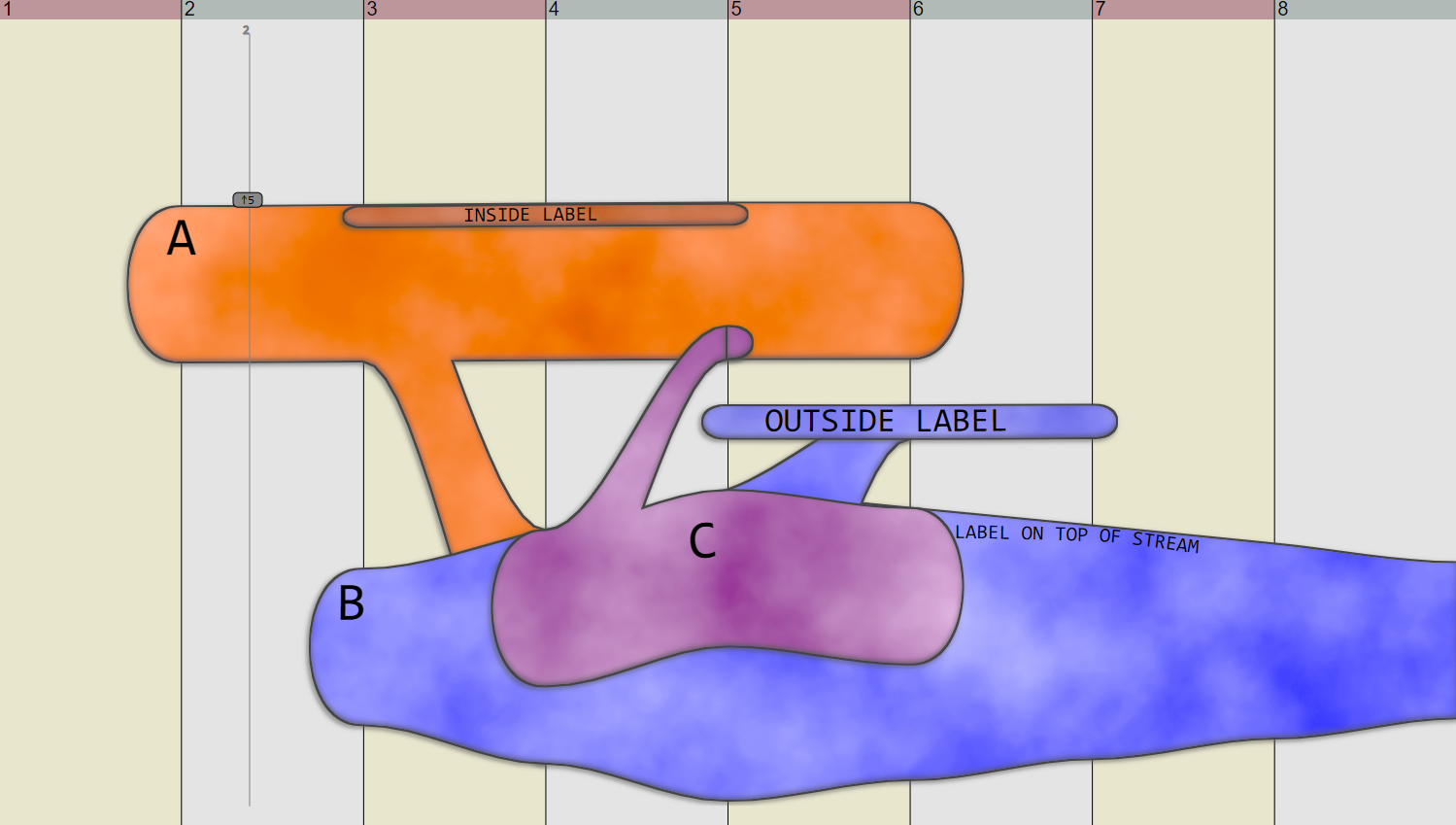}
    \caption{Visual result of the CSV data given in a. The force layout found a good separation of nodes and the artistic stylization was applied. Hovering over A shows the draggable button to change the node size at the given timepoint.}
    \label{fig:overview}
    \end{subfigure}
\caption{We defined 3 streams A, B, C, where C is nested inside B. B has a defined size of 10 at timepoint 5, and its size at other timepoints is calculated by linear interpolation between 10 and the default value (5) at the start and end point. C has a link to A, which results in a small representation of C nested inside A. A has a merge link to B, which means that no such nested node is created. We define 3 labels of different types, resulting in different representations.
}
    \label{my label}
\end{figure}

\noindent\textbf{Graph Layout:} We transform our data into a directed acyclic graph (DAG), similar to approaches found in TextFlow~\cite{cui2011textflow} and StoryFlow~\cite{liu2013storyflow}, to benefit from existing graph drawing algorithms when creating a layout. For each stream, we create one node at each discrete point in time (e.g., one node per year) within the time interval it is defined in and connect consecutive nodes via edges. For each label, we create a node at the referred to point in time, as well as an equal amount of nodes to the left and right, based on the length of the text. This representation of labels by several nodes enables the bending of text along the time axis. For outside labels we add an additional edge to a stream node.

We build on top of the SplitStreams~\cite{bolte2020splitstreams} approach to support a nested graph model, where every node can contain multiple other nodes, as well as splits and merges of streams. We utilize the same nesting mechanism to support labels inside and on top of streams.
Links are represented by an edge between two nodes, which must refer to different timepoints to satisfy the DAG constraints. If links span over several points in time, we generate intermediate link nodes as described by Sugiyama et al.~\cite{sugiyama1981methods}.

We utilize a force layout to generate the organic appearance of streams. Each node is fixed in the time dimension and its position can only vary in the second dimension based on the applied forces. Initially, all nodes are stacked above each other in the same order as defined in the data. A gravitational force applied to all nodes pushes them towards the center of the picture. A repulsive force between all nodes increases the spacing between individual streams and labels. We make sure that nodes stay inside the picture by introducing a collision detection at the border. A similar constraint is applied for nested nodes, so that they can not escape their parent elements. For simplification, collisions are not computed for actual stream shapes, but based on the individual nodes in each timestep.

We introduce different forces along different edge types for a fine-grained layout control. Edges between nodes of the same stream feature rather strong forces to keep the streams as straight as possible and to reduce stream crossings. These forces can be loosened to increase the wiggle of lines and thereby change the general stream appearance. Edges between labels and streams make sure that labels are drawn in close proximity to their connected stream. Finally, links between different entities keep merged, split, and connected streams closer together. By modifying their force, we can control how likely a stream is to bend for such a connection.

\noindent\textbf{Artistic Style:} We draw Cubic Bezier curves between connected nodes of the graph. Each stream is given a solid color, on top of which we apply image processing techniques to acquire the desired style. We multiply the colored stream with a grey scale fractal noise to integrate irregularities. The colors appear more hand-drawn than the solid color that is free from any imperfections. We add a strong black outline and both an inner and an outer shadow to each element of the chart. The latter are achieved by multiplying each element's pixel image with a black and offset copy of itself.

\begin{figure*}
	\centering{\includegraphics[width=0.9\textwidth]{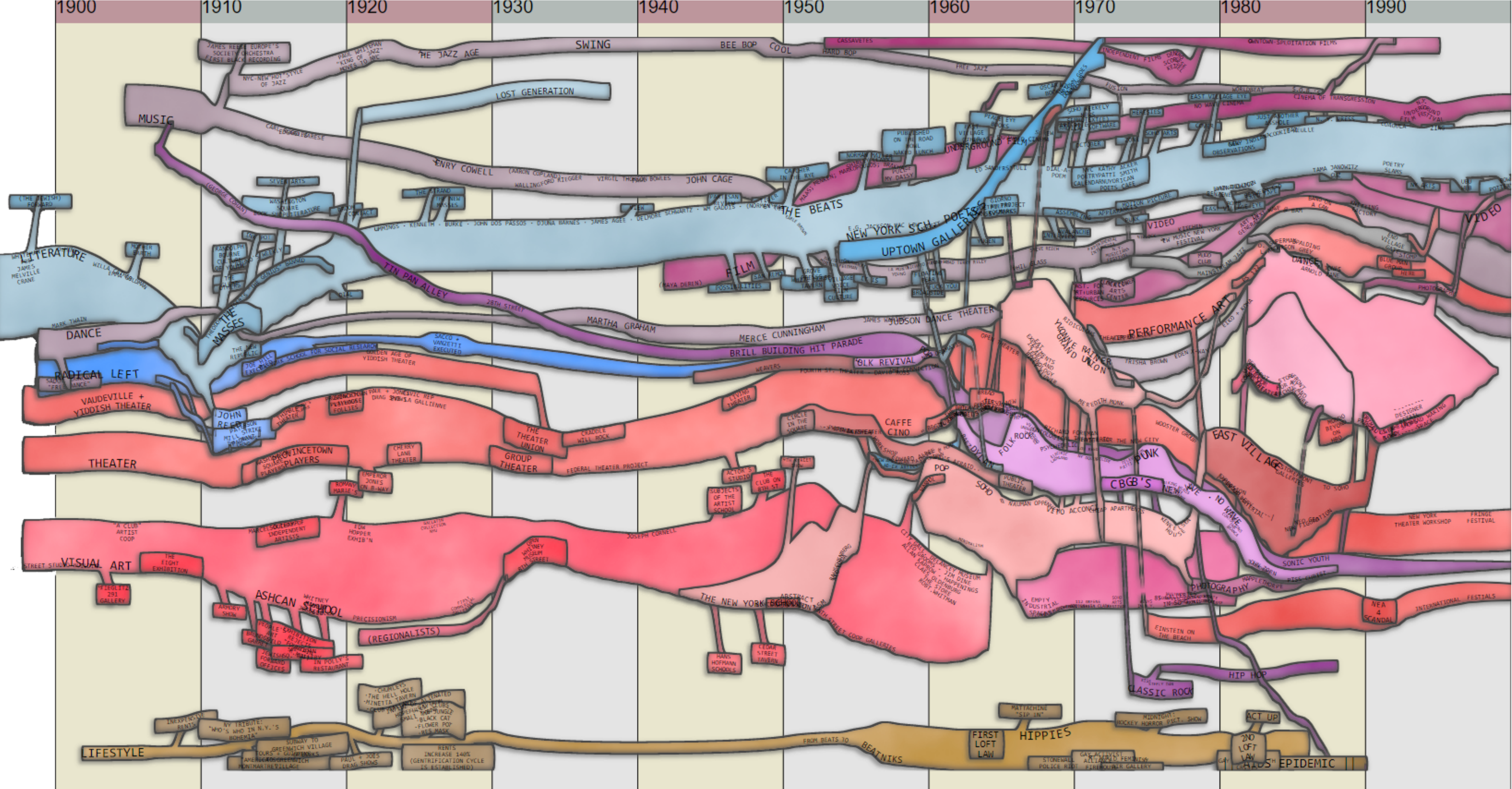}}
	\caption{Digital recreation of Ward Shelley's Downtown Body, including 44 streams, 61 links, and 369 labels.}
	\label{fig:result}
\end{figure*}

\noindent\textbf{Interactions:} The manual creation of CSV data for the generation of organic narrative charts might be tedious. We therefore designed several interaction techniques that we consider to be crucial for the development of an authoring tool for artists.


By clicking and holding the mouse button (or touch) in an empty area and dragging it over the chart, we can define a start and end point for a new stream. The same drag\&drop interaction can be utilized to create a link between two entities. If only one of the two points has an underlying entity and the other lies in an empty area, this interaction will create a new stream and a link connecting both entities. The user can define if the created connection is supposed to merge (or split) the connected stream, or if the new stream will be nested at the connection point.

When hovering over a stream, a draggable button will be shown that allows for the adjustment of the stream size at that point in time. In \autoref{fig:overview} we hover over stream A between timesteps 2 and 3. The size of each stream node is calculated through linear interpolation between consecutive size definitions and a default value at the start and end point.
Clicking at a stream will create a label at that point in time and a popup window enables the user to enter the desired text and type for the label.

\noindent\textbf{Implementation:} Our prototype implementation is based on D3~\cite{Bostock2011} and standard web technologies  (JavaScript ES6, HTML5, CSS3). Additionally to the interactions mentioned, we allow for the direct manipulation of the CSV data for detailed changes and update the chart on every change. We utilize the D3-integrated force layout and draw streams between connected nodes through the SplitStreams~\cite{bolte2020splitstreams} library. New nodes are added into the existing graph and the force layout runs based on the previous node positions to minimize calculation times.
We utilize SVG filters as image processing techniques to apply the artistic style. The output is in the standard SVG format, which allows for further processing in common vector graphics tools.
The result of our technique applied to a manually created data set mirroring the data in the \textit{Downtown Body} painting can be seen in \autoref{fig:result}.

\section{Discussion and Limitations}\label{sec:discussion}

The creation of our prototype yielded results of promising visual quality to improve existing narrative charts and added an artistic appeal that is normally only achieved by hand-drawn diagrams.
One of the main problems we face in the creation process is the correct definition of parameters for the force layout to properly separate individual nodes. If the authoring tool is meant to be used by the general public, they should not be exposed to a manual control of these parameters, but instead be given a fully-automated, or at least largely simplified, control sequence. This is especially true for the positioning of labels, which might not be readable when curves bend too strong or overlaps occur. 

We currently do not provide means for the integration of images for visual embellishment, and can not bend the time axis as featured in some of Ward Shelley's works. We can not capture the thematic content of the diagram to create a bigger picture like artists can convey. We therefore see our application more as a planning tool for artists to simplify the creation of such charts, as well as an option to easily create visually appealing diagrams from digital data.

\section{Conclusion}\label{sec:conclusion}

We presented organic narrative charts, a digital recreation of Ward Shelley's diagrammatic paintings from data, that is capable of conveying the complex relationships of entities in his work and provides a similar visual appeal. We implemented a prototype of an authoring tool and compared the digital results to the original work.
We provide the full source code at \textit{https://github.com/cadanox/orcha}.



\nocite{*}

\bibliographystyle{eg-alpha-doi}
\bibliography{eurovis2020-OrCha}


\end{document}